%% file: mathur.tex
\documentclass{svmult}
\usepackage{epsfig,graphicx} 
\usepackage[bottom]{footmisc}
\usepackage{natbib,url}      
\bibpunct{(}{)}{;}{a}{}{,}   
\def\cite#1{\citealp{#1}}    
\def\authorindex#1{}  
\def\figspath{.}  

\begin{document}\newcount\preprintheader\preprintheader=1


\input{rr-assp-defs}


\title*{Low-Degree High-Frequency $p$ and $g$ Modes in the Solar Core}

\titlerunning{Dynamics of solar core: $p$ and $g$ modes}

\author{S. Mathur\inst{1}
	\and
	R.~A. Garc\'ia\inst{2}
	\and
	A. Eff-Darwich\inst{3,4}}

\authorindex{Mathur, S.} 
\authorindex{Garc\'ia, R. A.}
\authorindex{Eff-Darwich, A.}

\authorrunning{Mathur et al.}

\institute{Indian Institute of Astrophysics, Bangalore, India
		\and
		Laboratoire AIM, CEA/DSM-CNRS-Univ. Paris Diderot; Gif-sur-Yvette, France
		\and
		Departamento de Edafolog\'ia y Geolog\'ia, Univ. de La Laguna, Tenerife, Spain
		\and 
		Instituto de Astrof\'isica de Canarias, Tenerife, Spain}

\maketitle

\setcounter{footnote}{0}  

\begin{abstract} 
    Solar gravity ($g$) modes propagate within the radiative part of
    the solar interior and are highly sensitive to the physical
    conditions of the solar core.  They would represent the best tool
    to infer the structure and dynamics of the radiative interior, in
    particular the core, if they were properly detected and
    characterized. Although individual rotational splittings for $g$
    modes have not yet been calculated, we have to understand the
    effect of these modes, and also low-degree high-frequency $p$
    modes, on the inversion of the solar rotation rate between 0.1 and
    0.2~R$_\odot$. In this work, we follow the methodology developed
    in Mathur et al.\ (2008) and \citet{2008AN....329..476G}, adding
    $g$ modes and low-degree high-frequency $p$ modes to artificial
    inversion data sets, in order to study how they convey
    information on the solar core rotation.
\end{abstract}

\section{Introduction}      \label{yourname-sec:introduction}

Helioseismology has improved our understanding of the dynamical
processes occurring within the solar interior \citep[ and references
therein]{ThoJCD2003}, in particular the convective zone and the
radiative zone down to 0.3~R$_\odot$ \citep{CouGar2003,
GarCor2004}. However, the solar deep interior is still poorly
constrained and the possible effect of the rotation rate in these
regions on the solar structure distribution (\eg\ sound speed and/or
density) remains uncertain \citep{2004A&A...425..229M,
2005A&A...440..653M, 2008JPhCS.118a2030T}. These uncertainties should
be added to the recently found discrepancies between helioseismic
observations and theoretical modelling
\citep{2005ApJ...620L.129A,STCCou2004,2007ApJ...668..594M,2007A&A...469.1145Z}
when the new surface chemical abundances of
\citet{2005ASPC..336...25A} are used to calculate the structural model
of the solar interior. In this context, gravity modes would add strong
observational constraints about the structure and dynamics of the
solar core. A few individual candidates have been detected
\citep{STCGar2004}, but also global asymptotic properties
\citep{2007Sci...316.1591G, 2008AN....329..476G} that were used by
\citet{2008SoPh..251..135G} to infer the rotation rate of the solar
core and to constrain various physical quantities.

\begin{table*}[ht]
  \begin{center} 
\caption{Artificial data sets, with $\nu$ (mHz) the mode frequency 
and $\epsilon$ its uncertainty.}
\begin{tabular}[h]{lc|c|c|cc}
\hline 
Data set &  $g$ modes  &  $p$ modes $\ell=1, 3$  &  $\epsilon$ for $\ell=1, 3$ & $p$ modes  $\ell > 3$ \\
\hline
  Set {1} &  -  &  $1\le\nu\le2.3  $ & $\epsilon$& $1\le\nu\le3.9$  \\
  Set {2} &  -  &  $1\le\nu\le3.9  $ & $\epsilon$& $1\le\nu\le3.9$  \\
  Set {3} &  -  &  $1\le\nu\le3.9  $ & $\epsilon$/3 & $1\le\nu\le3.9$  \\
  Set {4} &  -  &  $1\le\nu\le3.9  $ & $\epsilon$/6 & $1\le\nu\le3.9$  \\
 Set {5} &   $\ell=2$, $n=-3$  & $1\le\nu\le3.9$ & $\epsilon$& $1\le\nu\le3.9$  \\
Set {6} & $\ell=1-2$, $n=-1$ to $-10$ &  $1\le\nu\le3.9$ & $\epsilon$& $1\le\nu\le3.9$ \\
\hline
     \end{tabular}
    \label{mathur-tab:sets}
  \end{center}
\end{table*}

An artificial rotation rate (blue curve in Fig.~\ref{mathur-fig:001})
was used to calculate the frequency splittings for $p$ and $g$
modes. The shape of this profile (note that between 0.15 and
0.2~R$_\odot$ the rate is 3 times that in the rest of the
radiative zone) was choosen to study the effect of the different modes
used in the inversion sets (Table~\ref{mathur-tab:sets}). The
inversions were carried out through the Regularised Least Squared
methodology
\citep{Eff-Darwich1997}, the mode sets came from the artificial
rotational rate, and the data uncertainties and noise were
calculated from observations \citep{2008SoPh..251..119G, Kor2005}. We
know that the inner turning points of low-degree high-frequency $p$
modes lay around 0.1-0.2~R$_\odot$ \citep{2008SoPh..251..119G},
whereas those of $g$ modes are confined in a small region below
0.2~R$_\odot$ \citep{2008A&A...484..517M}. A priori, these modes would
significantly improve the inversion results in the solar core, but it
is necessary to understand the effect of observational uncertainties
and the number of modes to set the accuracy needed from peak fitting
techniques to properly characterize the core.


\section{The effect of low-degree high-frequency $p$ modes}

\begin{figure}  
  \centering
  \begin{tabular}{p{5.5cm}p{5.5cm}}
  	\includegraphics[width=5.5cm]{\figspath/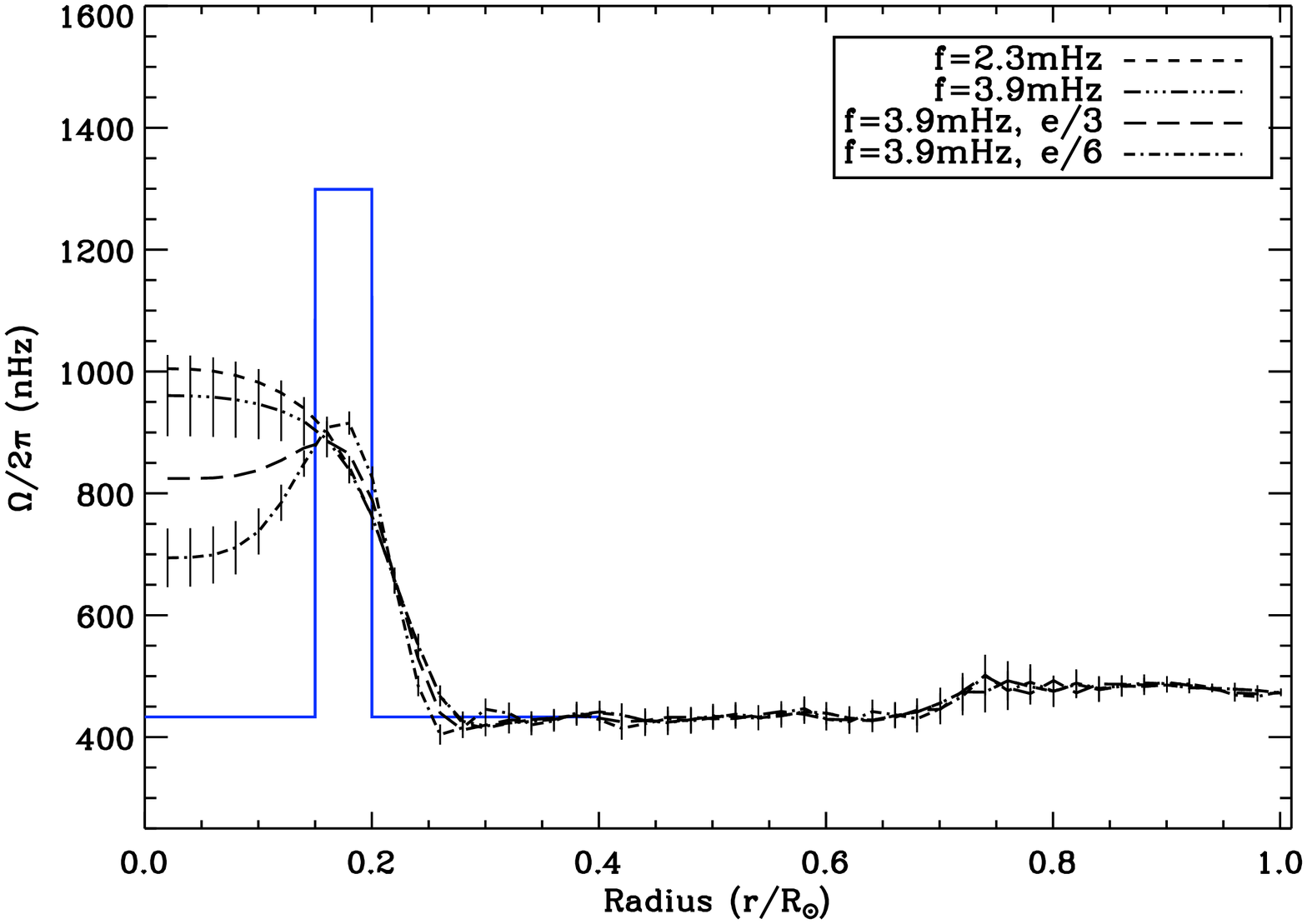}&
	\includegraphics[width=5.5cm]{\figspath/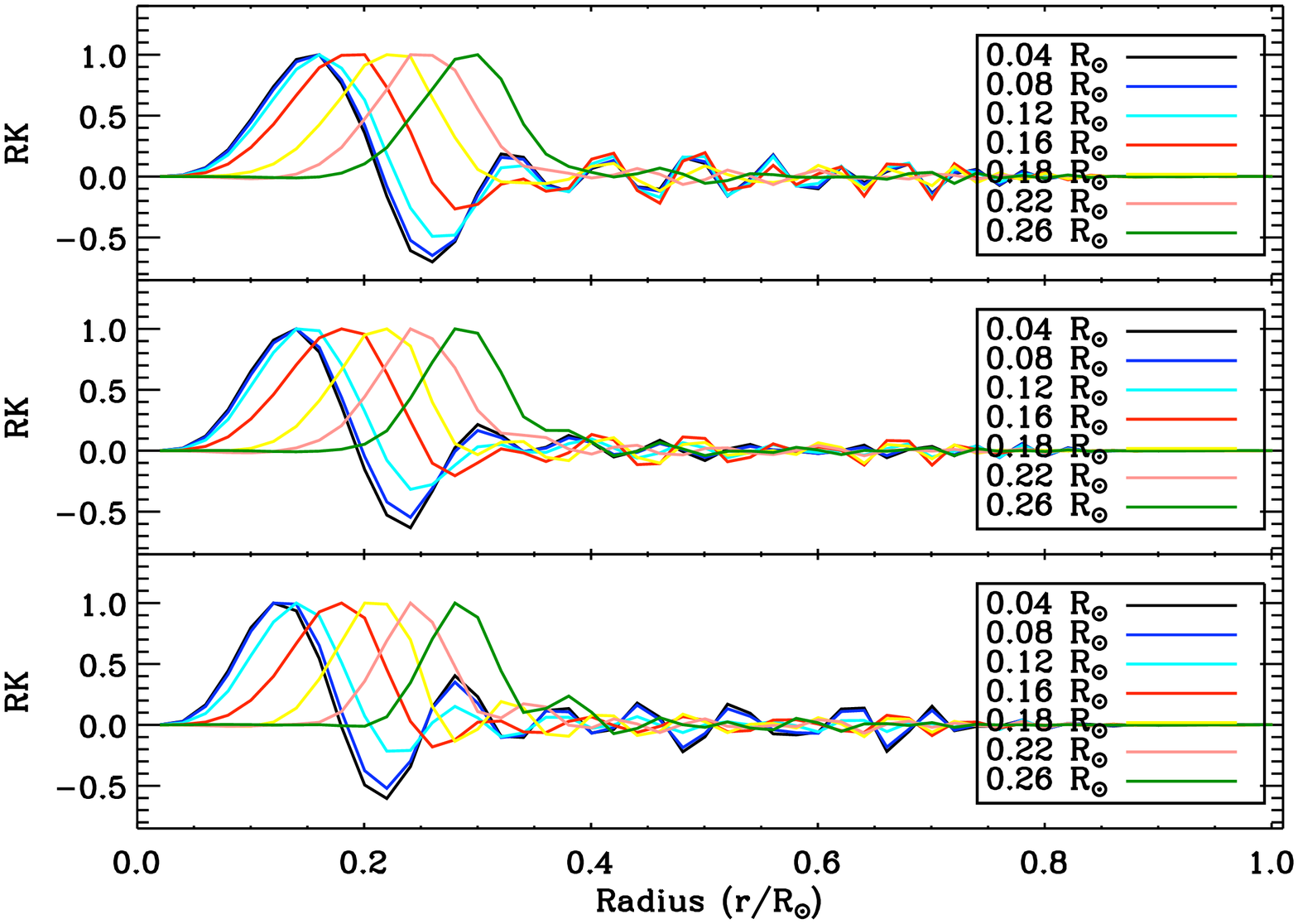}
	\end{tabular}
  
  \caption[]{\label{mathur-fig:001}
%
Left: Rotation profiles (equatorial) reconstructed from different sets
of $p$ modes. Right: Resolution kernels of sets 1 (top), 3 (middle),
and 4 (bottom).
}\end{figure}

When low-degree high-frequency $p$ modes are added to the inversion
data sets (1 to 4), the increase in the rotation rate below
0.2~R$_\odot$ is detected in all cases but the maximum rate is
not recovered (Fig.~\ref{mathur-fig:001}). Only for
data sets 3 and 4, with the error bars significantly lowered, is the
decrease of the rotation rate in the solar coredetected, although
the actual values are not recovered. The resolution kernels show that
the deepest point sensed by the inversion shifts from 0.15
down to 0.12~R$_\odot$, explaining the peak sensitivity of these
sets to the region between 0.1 and 0.15~R$_\odot$.

\section{The effect of $g$ modes in the inversions}

\begin{figure}  
  \centering
  \begin{tabular}{p{5.5cm}p{5.5cm}}
  	\includegraphics[width=5.5cm]{\figspath/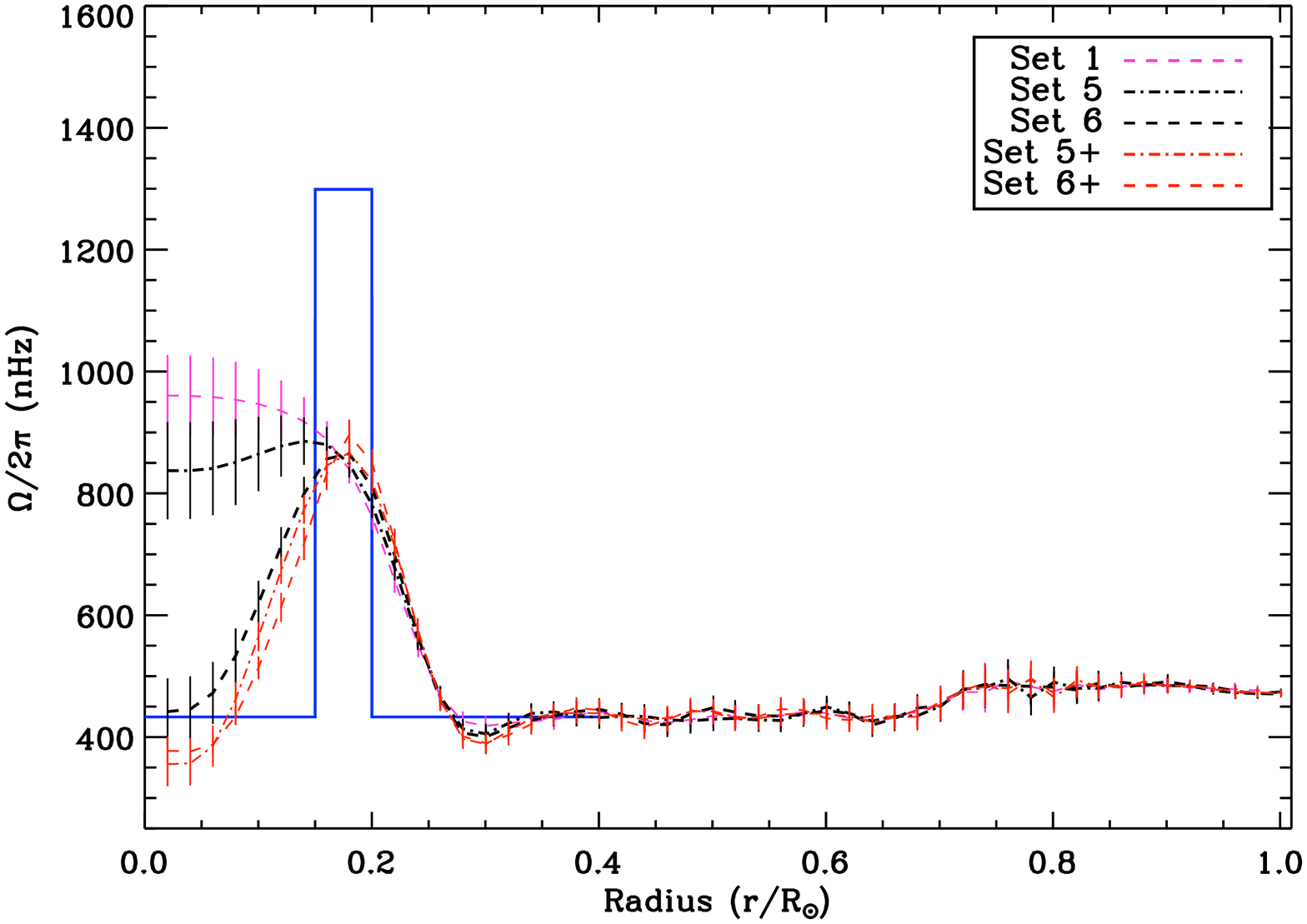}&
	\includegraphics[width=5.5cm]{\figspath/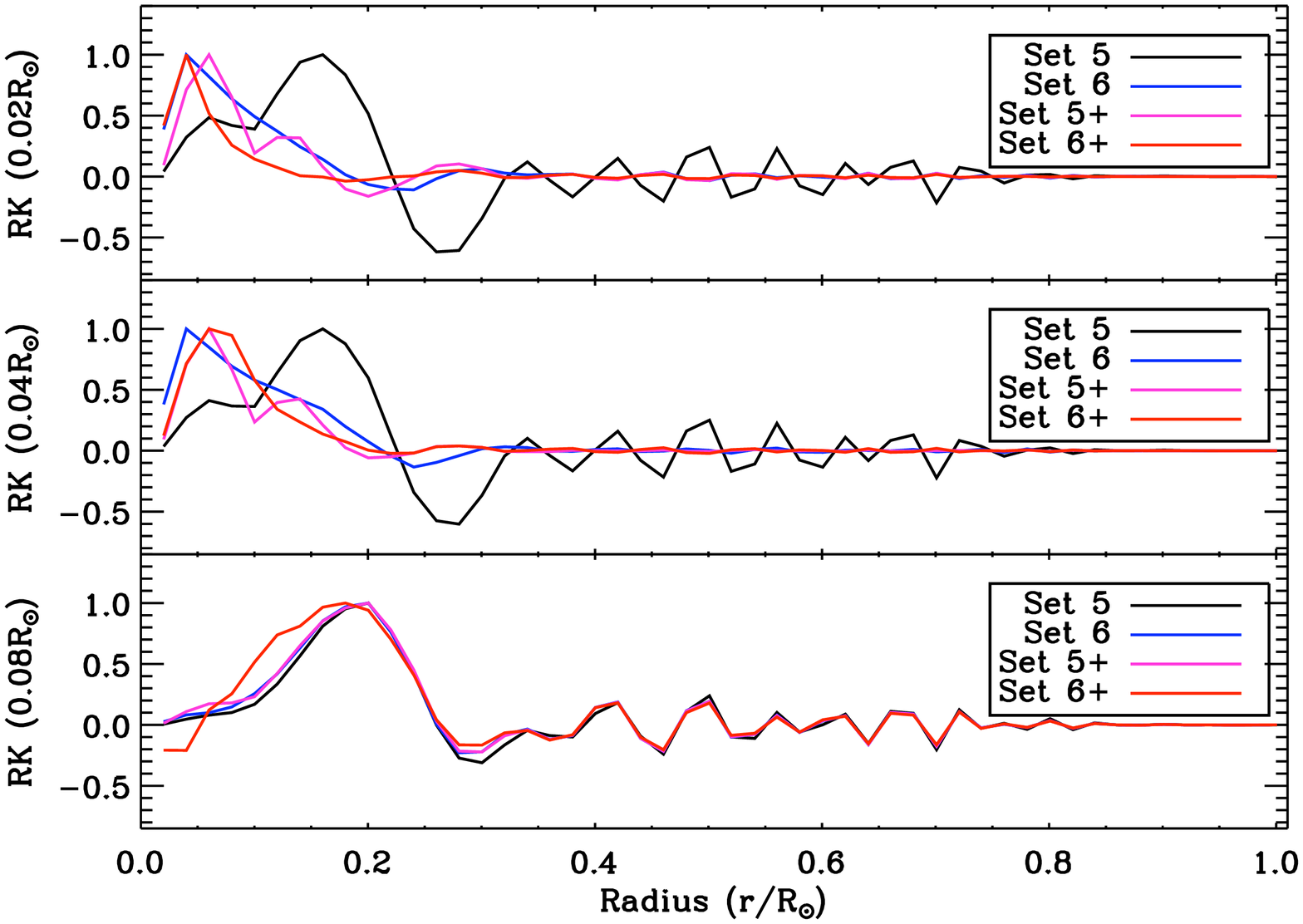}\\

	\end{tabular}
  \caption[]{\label{mathur-fig:002}
%
Left: Same as Fig.~\ref{mathur-fig:001} with 1 and 20 $g$ modes
($\epsilon$=75 and 10 nHz). Right: Resolution kernels for sets 5, 6,
5+ and 6+ at 0.02, 0.04, and 0.08~R$_\odot$.
}\end{figure}

The effect of adding $g$ modes on the inversions was analyzed adding 1
mode (set 5) and 20 modes (set 6) $g$ to set 2
(Fig.~\ref{mathur-fig:002}). In both cases, the uncertainties in the
$g$ modes were 75~nHz, although two more sets (set 5+ and 6+) with the
uncertainties reduced to 10~nHz were also computed. The inversion of
set 5 is comparable to that of set 3 where no $g$ modes were
used. This result reflects the fact that individual $g$ modes should
be detected with accuracies significantly lower than 75~nHz. For sets
6, 5+ and 6+, the rotation rate recovered in the core is similar to
the actual artificial rate. Hence, the detection of a few very
accurate $g$ modes will give similar information as a larger set of
less accurate data. The resolution kernels show that the deepest point
at which the inversion is sensitive shifts from 0.15 down to
0.03~R$_\odot$. However, the sensitivity is still quite low between
0.05 and 0.15~R$_\odot$ (set 6 in Fig.~\ref{mathur-fig:002}) when only
$g$ modes are used in the inversion.  With a lower error bar, we probe
several points between 0.03 and 0.15~R$_\odot$. Moreover, the
resolution kernels are quite broad, explaining that we do not retrieve
the rotation rate between 0.15 and 0.2~R$_\odot$ (average on the width
of these kernels). To study this region it would be necessary to
significantly reduce the observational uncertainties for low degree
and high frequency $p$ modes.  Finally, resolution kernels for sets 6,
5+ and 6+, exhibit less wiggles above $\approx$0.3~R$_\odot$: they are
less polluted by the rotation of the outer layers.

\section{Conclusion}                   \label{yourname-sec:conclusion}

Gravity modes are necessary to improve our inferences on the
rotational rate at the solar core. However, these modes should reach a
certain level of accuracy, namely uncertainties for the frequency
splittings significantly lower than 75~nHz, to assure that at least
one $g$ mode will be sufficient to greatly improve our sensitivity to
the dynamics of the core. Although gravity modes will certainly
improve inversion results below 0.1~R$_\odot$, it would be necessary
to properly characterize low-degree and high-frequency $p$ modes to
resolve the region between 0.1 - 0.2~R$_\odot$.


\begin{small}






\end{small}

\end{document}

%% file: rr-assp-defs.tex

\def\thisvolume{these proceedings}

\def\aj{{AJ}}			
\def\araa{{ARA\&A}}		
\def\apj{{ApJ}}			
\def\apjl{{ApJ}}		
\def\apjs{{ApJS}}		
\def\ao{{Appl.\ Optics}} 
\def\apss{{Ap\&SS}}		
\def\aap{{A\&A}}		
\def\aapr{{A\&A~Rev.}}		
\def\aaps{{A\&AS}}		
\def\an{{Astron.\ Nachrichten}}
\def\aspcs{{ASP Conf.\ Ser.}}
\def\assp{{Astrophys.\ \& Space Sci.\ Procs., Springer, Heidelberg}}
\def\azh{{AZh}}			
\def\baas{{BAAS}}		
\def\jrasc{{JRASC}}	
\def\memras{{MmRAS}}		
\def\mnras{{MNRAS}}
\def\nat{{Nat}}		
\def\pra{{Phys.\ Rev.\ A}} 
\def\prb{{Phys.\ Rev.\ B}}		
\def\prc{{Phys.\ Rev.\ C}}		
\def\prd{{Phys.\ Rev.\ D}}		
\def\prl{{Phys.\ Rev.\ Lett.}} 
\def\pasp{{PASP}}
\def\pasj{{PASJ}}		
\def\qjras{{QJRAS}}
\def\science{{Sci}}		
\def\skytel{{S\&T}}		
\def\solphys{{Solar\ Phys.}} 
\def\sovast{{Soviet\ Ast.}}  
\def\ssr{{Space\ Sci.\ Rev.}}
\def\svassp{{Astrophys.\ Space Sci.\ Procs., Springer, Heidelberg}}
\def\zap{{ZAp}}			
\let\astap=\aap
\let\apjlett=\apjl
\let\apjsupp=\apjs
\def\grl{{Geophys.\ Res.\ Lett.}}  
\def\jgr{{J. Geophys.\ Res.}} 

\def\ion#1#2{{\rm #1}\,{\uppercase{#2}}}  
\def\deg{\hbox{$^\circ$}}
\def\sun{\hbox{$\odot$}}
\def\earth{\hbox{$\oplus$}}
\def\la{\mathrel{\hbox{\rlap{\hbox{\lower4pt\hbox{$\sim$}}}\hbox{$<$}}}}
\def\ga{\mathrel{\hbox{\rlap{\hbox{\lower4pt\hbox{$\sim$}}}\hbox{$>$}}}}
\def\sq{\hbox{\rlap{$\sqcap$}$\sqcup$}}
\def\arcmin{\hbox{$^\prime$}}
\def\arcsec{\hbox{$^{\prime\prime}$}}
\def\fd{\hbox{$.\!\!^{\rm d}$}}
\def\fh{\hbox{$.\!\!^{\rm h}$}}
\def\fm{\hbox{$.\!\!^{\rm m}$}}
\def\fs{\hbox{$.\!\!^{\rm s}$}}
\def\fdg{\hbox{$.\!\!^\circ$}}
\def\farcm{\hbox{$.\mkern-4mu^\prime$}}
\def\farcs{\hbox{$.\!\!^{\prime\prime}$}}
\def\fp{\hbox{$.\!\!^{\scriptscriptstyle\rm p}$}}
\def\micron{\hbox{$\mu$m}}
\def\onehalf{\hbox{$\,^1\!/_2$}}	
\def\onethird{\hbox{$\,^1\!/_3$}}
\def\twothirds{\hbox{$\,^2\!/_3$}}
\def\onequarter{\hbox{$\,^1\!/_4$}}
\def\threequarters{\hbox{$\,^3\!/_4$}}
\def\ubv{\hbox{$U\!BV$}}		
\def\ubvr{\hbox{$U\!BV\!R$}}		
\def\ubvri{\hbox{$U\!BV\!RI$}}		
\def\ubvrij{\hbox{$U\!BV\!RI\!J$}}		
\def\ubvrijh{\hbox{$U\!BV\!RI\!J\!H$}}		
\def\ubvrijhk{\hbox{$U\!BV\!RI\!J\!H\!K$}}		
\def\ub{\hbox{$U\!-\!B$}}		
\def\bv{\hbox{$B\!-\!V$}}		
\def\vr{\hbox{$V\!-\!R$}}		
\def\ur{\hbox{$U\!-\!R$}}


\def\labelitemi{{\bf --}}  

\def\rmit#1{{\it #1}}              
\def\rmit#1{{\rm #1}}              
\def\etal{\rmit{et al.}}           
\def\etc{\rmit{etc.}}           
\def\ie{\rmit{i.e.,}}              
\def\eg{\rmit{e.g.,}}              
\def\cf{cf.}                       
\def\viz{\rmit{viz.}}
\def\vs{\rmit{vs.}}

\def\rot{\hbox{\rm rot}}
\def\div{\hbox{\rm div}}
\def\lesssim{\mathrel{\hbox{\rlap{\hbox{\lower4pt\hbox{$\sim$}}}\hbox{$<$}}}}
\def\gtrsim{\mathrel{\hbox{\rlap{\hbox{\lower4pt\hbox{$\sim$}}}\hbox{$>$}}}}
\def\dif{\: {\rm d}}                       
\def\ep{\:{\rm e}^}                        
\def\dash{\hbox{$\,-\,$}}                  
\def\is{\!=\!}                             

\def\starname#1#2{${#1}$\,{\rm {#2}}}  
\def\Teff{\hbox{$T_{\rm eff}$}}   

\def\kms{\hbox{km$\;$s$^{-1}$}}
\def\ms{\hbox{m$\;$s$^{-1}$}}
\def\Mxcm{\hbox{Mx\,cm$^{-2}$}}    

\def\Bapp{\hbox{$B_{\rm app}$}}    

\def\komega{($k, \omega$)}                 
\def\kf{($k_h,f$)}                         
\def\VminI{\hbox{$V\!\!-\!\!I$}}           
\def\IminI{\hbox{$I\!\!-\!\!I$}}           
\def\VminV{\hbox{$V\!\!-\!\!V$}}           
\def\Xt{\hbox{$X\!\!-\!t$}}                

\def\level #1 #2#3#4{$#1 \: ^{#2} \mbox{#3} ^{#4}$}   

\def\specchar#1{\uppercase{#1}}    
\def\AlI{\mbox{Al\,\specchar{i}}}  
\def\BI{\mbox{B\,\specchar{i}}} 
\def\BII{\mbox{B\,\specchar{ii}}}  
\def\BaI{\mbox{Ba\,\specchar{i}}}  
\def\BaII{\mbox{Ba\,\specchar{ii}}} 
\def\CI{\mbox{C\,\specchar{i}}} 
\def\CII{\mbox{C\,\specchar{ii}}} 
\def\CIII{\mbox{C\,\specchar{iii}}} 
\def\CIV{\mbox{C\,\specchar{iv}}} 
\def\CaI{\mbox{Ca\,\specchar{i}}} 
\def\CaII{\mbox{Ca\,\specchar{ii}}} 
\def\CaIII{\mbox{Ca\,\specchar{iii}}} 
\def\CoI{\mbox{Co\,\specchar{i}}} 
\def\CrI{\mbox{Cr\,\specchar{i}}} 
\def\CriI{\mbox{Cr\,\specchar{ii}}} 
\def\CsI{\mbox{Cs\,\specchar{i}}} 
\def\CsII{\mbox{Cs\,\specchar{ii}}} 
\def\CuI{\mbox{Cu\,\specchar{i}}} 
\def\FeI{\mbox{Fe\,\specchar{i}}} 
\def\FeII{\mbox{Fe\,\specchar{ii}}} 
\def\FeIX{\mbox{Fe\,\specchar{ix}}}
\def\FeX{\mbox{Fe\,\specchar{x}}}
\def\FeXVI{\mbox{Fe\,\specchar{xvi}}}
\def\FrI{\mbox{Fr\,\specchar{i}}}
\def\HI{\mbox{H\,\specchar{i}}} 
\def\HII{\mbox{H\,\specchar{ii}}} 
\def\Hmin{\hbox{\rmH$^{^{_{\scriptstyle -}}}$}}      
\def\Hemin{\hbox{{\rm He}$^{^{_{\scriptstyle -}}}$}} 
\def\HeI{\mbox{He\,\specchar{i}}} 
\def\HeII{\mbox{He\,\specchar{ii}}} 
\def\HeIII{\mbox{He\,\specchar{iii}}} 
\def\KI{\mbox{K\,\specchar{i}}} 
\def\KII{\mbox{K\,\specchar{ii}}} 
\def\KIII{\mbox{K\,\specchar{iii}}} 
\def\LiI{\mbox{Li\,\specchar{i}}} 
\def\LiII{\mbox{Li\,\specchar{ii}}} 
\def\LiIII{\mbox{Li\,\specchar{iii}}} 
\def\MgI{\mbox{Mg\,\specchar{i}}} 
\def\MgII{\mbox{Mg\,\specchar{ii}}} 
\def\MgIII{\mbox{Mg\,\specchar{iii}}} 
\def\MnI{\mbox{Mn\,\specchar{i}}} 
\def\NI{\mbox{N\,\specchar{i}}}
\def\NIV{\mbox{N\,\specchar{iv}}}
\def\NaI{\mbox{Na\,\specchar{i}}}
\def\NaII{\mbox{Na\,\specchar{ii}}}
\def\NaIII{\mbox{Na\,\specchar{iii}}}
\def\NeVIII{\mbox{Ne\,\specchar{viii}}} 
\def\NiI{\mbox{Ni\,\specchar{i}}} 
\def\NiII{\mbox{Ni\,\specchar{ii}}}
\def\NiIII{\mbox{Ni\,\specchar{iii}}} 
\def\OI{\mbox{O\,\specchar{i}}} 
\def\OVI{\mbox{O\,\specchar{vi}}}
\def\RbI{\mbox{Rb\,\specchar{i}}} 
\def\SII{\mbox{S\,\specchar{ii}}} 
\def\SiI{\mbox{Si\,\specchar{i}}} 
\def\SiII{\mbox{Si\,\specchar{ii}}} 
\def\SrI{\mbox{Sr\,\specchar{i}}}
\def\SrII{\mbox{Sr\,\specchar{ii}}}
\def\TiI{\mbox{Ti\,\specchar{i}}} 
\def\TiII{\mbox{Ti\,\specchar{ii}}} 
\def\TiIII{\mbox{Ti\,\specchar{iii}}} 
\def\TiIV{\mbox{Ti\,\specchar{iv}}} 
\def\VI{\mbox{V\,\specchar{i}}} 
\def\HtwoO{\mbox{H$_2$O}}        
\def\Otwo{\mbox{O$_2$}}          

\def\Halpha{\mbox{H\hspace{0.1ex}$\alpha$}} 
\def\Ha{\mbox{H\hspace{0.2ex}$\alpha$}}
\def\Hbeta{\mbox{H\hspace{0.2ex}$\beta$}}
\def\Hgamma{\mbox{H\hspace{0.2ex}$\gamma$}}
\def\Hdelta{\mbox{H\hspace{0.2ex}$\delta$}}
\def\Hepsilon{\mbox{H\hspace{0.2ex}$\epsilon$}}
\def\Hzeta{\mbox{H\hspace{0.2ex}$\zeta$}}
\def\Lyalpha{\mbox{Ly$\hspace{0.2ex}\alpha$}}
\def\Lybeta{\mbox{Ly$\hspace{0.2ex}\beta$}}
\def\Lygamma{\mbox{Ly$\hspace{0.2ex}\gamma$}}
\def\Lycont{\mbox{Ly\hspace{0.2ex}{\small cont}}}
\def\Baalpha{\mbox{Ba$\hspace{0.2ex}\alpha$}}
\def\Babeta{\mbox{Ba$\hspace{0.2ex}\beta$}}
\def\Bacont{\mbox{Ba\hspace{0.2ex}{\small cont}}}
\def\Paalpha{\mbox{Pa$\hspace{0.2ex}\alpha$}}
\def\Bralpha{\mbox{Br$\hspace{0.2ex}\alpha$}}

\def\NaD{\mbox{Na\,\specchar{i}\,D}}    
\def\NaDone{\mbox{Na\,\specchar{i}\,\,D$_1$}}
\def\NaDtwo{\mbox{Na\,\specchar{i}\,\,D$_2$}}
\def\NaID{\mbox{Na\,\specchar{i}\,\,D}}
\def\NaIDone{\mbox{Na\,\specchar{i}\,\,D$_1$}}
\def\NaIDtwo{\mbox{Na\,\specchar{i}\,\,D$_2$}}
\def\Done{\mbox{D$_1$}}
\def\Dtwo{\mbox{D$_2$}}

\def\Mgbone{\mbox{Mg\,\specchar{i}\,b$_1$}}
\def\Mgbtwo{\mbox{Mg\,\specchar{i}\,b$_2$}}
\def\Mgbthree{\mbox{Mg\,\specchar{i}\,b$_3$}}
\def\MgIb{\mbox{Mg\,\specchar{i}\,b}}
\def\MgIbone{\mbox{Mg\,\specchar{i}\,b$_1$}}
\def\MgIbtwo{\mbox{Mg\,\specchar{i}\,b$_2$}}
\def\MgIbthree{\mbox{Mg\,\specchar{i}\,b$_3$}}

\def\CaIIK{\mbox{Ca\,\specchar{ii}\,K}}       
\def\CaIIH{\mbox{Ca\,\specchar{ii}\,H}}
\def\CaIIHK{\mbox{Ca\,\specchar{ii}\,H\,\&\,K}}
\def\HK{\mbox{H\,\&\,K}}
\def\Kthree{\mbox{K$_3$}}      
\def\Hthree{\mbox{H$_3$}}
\def\Ktwo{\mbox{K$_2$}}
\def\Htwo{\mbox{H$_2$}}
\def\Kone{\mbox{K$_1$}}     
\def\Hone{\mbox{H$_1$}}     
\def\KtwoV{\mbox{K$_{2V}$}}
\def\KtwoR{\mbox{K$_{2R}$}}
\def\KoneV{\mbox{K$_{1V}$}}
\def\KoneR{\mbox{K$_{1R}$}}
\def\HtwoV{\mbox{H$_{2V}$}}
\def\HtwoR{\mbox{H$_{2R}$}}
\def\HoneV{\mbox{H$_{1V}$}}
\def\HoneR{\mbox{H$_{1R}$}}

\def\hk{\mbox{h\,\&\,k}}
\def\kthree{\mbox{k$_3$}}    
\def\hthree{\mbox{h$_3$}}
\def\ktwo{\mbox{k$_2$}}
\def\htwo{\mbox{h$_2$}}
\def\kone{\mbox{k$_1$}}     
\def\hone{\mbox{h$_1$}}     
\def\ktwoV{\mbox{k$_{2V}$}}
\def\ktwoR{\mbox{k$_{2R}$}}
\def\koneV{\mbox{k$_{1V}$}}
\def\koneR{\mbox{k$_{1R}$}}
\def\htwoV{\mbox{h$_{2V}$}}
\def\htwoR{\mbox{h$_{2R}$}}
\def\honeV{\mbox{h$_{1V}$}}
\def\honeR{\mbox{h$_{1R}$}}

\ifnum\preprintheader=1     
\makeatletter  
\def\@maketitle{\newpage
\markboth{}{}%
  {\em \footnotesize To appear in ``Magnetic Coupling between the Interior 
       and the Atmosphere of the Sun'', eds. S.~S.~Hasan and R.~J.~Rutten, 
       Astrophysics and Space Science Proceedings, Springer-Verlag, 
       Heidelberg, Berlin, 2009.}\par
 \def\lastand{\ifnum\value{@inst}=2\relax
                 \unskip{} \andname\
              \else
                 \unskip \lastandname\
              \fi}%
 \def\and{\stepcounter{@auth}\relax
          \ifnum\value{@auth}=\value{@inst}%
             \lastand
          \else
             \unskip,
          \fi}%
  \raggedright
 {\Large \bfseries\boldmath
  \pretolerance=10000
  \let\\=\newline
  \raggedright
  \hyphenpenalty \@M
  \interlinepenalty \@M
  \if@numart
     \chap@hangfrom{}
  \else
     \chap@hangfrom{\thechapter\thechapterend\hskip\betweenumberspace}
  \fi
  \ignorespaces
  \@title \par}\vskip .8cm
\if!\@subtitle!\else {\large \bfseries\boldmath
  \vskip -.65cm
  \pretolerance=10000
  \@subtitle \par}\vskip .8cm\fi
 \setbox0=\vbox{\setcounter{@auth}{1}\def\and{\stepcounter{@auth}}%
 \def\thanks##1{}\@author}%
 \global\value{@inst}=\value{@auth}%
 \global\value{auco}=\value{@auth}%
 \setcounter{@auth}{1}%
{\lineskip .5em
\noindent\ignorespaces
\@author\vskip.35cm}
 {\small\institutename\par}
 \ifdim\pagetotal>157\p@
     \vskip 11\p@
 \else
     \@tempdima=168\p@\advance\@tempdima by-\pagetotal
     \vskip\@tempdima
 \fi
}
\makeatother     
\fi